\documentclass[sigconf, 9pt]{acmart}

\usepackage{amsthm} 

\usepackage{mathtools}

\usepackage{fixmath}

\usepackage{multirow}
\usepackage{tabularx}
\newcolumntype{L}{>{\RaggedRight\arraybackslash}X}
\newcolumntype{C}{>{\centering\arraybackslash}X}
\usepackage{dcolumn}
\usepackage{colortbl}

\usepackage{algorithm}
\usepackage{algorithmic}

\usepackage{url}

\usepackage{subcaption}

\usepackage{enumitem}

\AtBeginDocument{%
  \providecommand\BibTeX{{%
    \normalfont B\kern-0.5em{\scshape i\kern-0.25em b}\kern-0.8em\TeX}}}

\copyrightyear{2024}
\acmYear{2024}
\setcopyright{rightsretained}
\acmConference[GLSVLSI '24]{Great Lakes Symposium on VLSI 2024}{June 12--14, 2024}{Clearwater, FL, USA}
\acmBooktitle{Great Lakes Symposium on VLSI 2024 (GLSVLSI '24), June 12--14, 2024, Clearwater, FL, USA}
\acmDOI{10.1145/3649476.3658736}
\acmISBN{979-8-4007-0605-9/24/06}
\begin{document}
\title[\resizebox{4.5in}{!}{NeuroBlend: Towards Low-Power yet Accurate Neural Network-Based Inference Engine Blending Binary and Fixed-Point Convolutions}]{ NeuroBlend: Towards Low-Power yet Accurate Neural Network-Based Inference Engine Blending Binary and Fixed-Point Convolutions}


\author{Arash Fayyazi}
\affiliation{%
  \institution{University of Southern California}
  \city{Los Angeles}
  \state{California}
  \country{USA}
}
\email{fayyazi@usc.edu}

\author{Mahdi Nazemi}
\affiliation{%
  \institution{University of Southern California}
  \city{Los Angeles}
  \state{California}
  \country{USA}
}
\email{mnazemi@usc.edu}

\author{Arya Fayyazi}
\affiliation{%
  \institution{University of Southern California}
  \city{Los Angeles}
  \state{California}
  \country{USA}
}
\email{afayyazi@usc.edu}

\author{Massoud Pedram}
\affiliation{%
  \institution{University of Southern California}
  \city{Los Angeles}
  \state{California}
  \country{USA}
}
\email{pedram@usc.edu}

\renewcommand{\shortauthors}{Fayyazi, et al.}

\begin{abstract}
This paper introduces NeuroBlend, a novel neural network architecture featuring a unique building block known as the Blend module. This module incorporates binary and fixed-point convolutions in its main and skip paths, respectively. There is a judicious deployment of batch normalizations on both main and skip paths inside the Blend module and in between consecutive Blend modules.
Additionally, we present a compiler and hardware architecture designed to map NeuroBlend models onto FPGA devices, aiming to minimize inference latency while maintaining high accuracy.
Our NeuroBlend-20 (NeuroBlend-18) model, derived from ResNet-20 (ResNet-18) trained on CIFAR-10 (CIFAR-100), achieves 88.0\% (73.73\%) classification accuracy, outperforming state-of-the-art binary neural networks by 0.8\% (1.33\%), with an inference time of 0.38ms per image, 1.4x faster than previous FPGA implementation for BNNs. 
Similarly, our BlendMixer model for CIFAR-10 attains 90.6\% accuracy(1.59\% less than full precision MLPMixer), with a 3.5x reduction in model size compared to full precision MLPMixer.
Furthermore, leveraging DSP blocks for 48-bit bitwise logic operations enables low-power FPGA implementation, yielding a 2.5x reduction in power consumption.
\end{abstract}



\maketitle
\vspace{-4mm}
\section{Introduction}

Deep neural networks (DNNs) have surpassed the accuracy of conventional machine learning models in many challenging domains, including computer vision \cite{DBLP:conf/cvpr/HuangLMW17} and natural language processing (NLP) \cite{DBLP:conf/naacl/DevlinCLT19}. 
Recently, inspired by the successes in NLP, \textit{transformers} \cite{DBLP:journals/corr/VaswaniSPUJGKP17} are adopted by the computer vision community. Built with self-attention layers, multi-layer perceptrons (MLPs), and skip connections, transformers make numerous breakthroughs on visual tasks \cite{DBLP:conf/iccv/LiuL00W0LG21}. To reduce the transformer model complexity, MLPMixers \cite{DBLP:conf/nips/TolstikhinHKBZU21}, which replace the multi-head self-attention module in transformers with a two-layer spatial MLP, are introduced. 

%
%
Unfortunately, many DNN-based inference engines have a high latency cost and use enormous hardware resources, which, in turn, prevent their deployment in latency-critical applications, especially on resource-constrained platforms. 
The high latency and large hardware cost are due to the fact that practical, high-quality deep learning models entail billions of arithmetic operations and millions of parameters, which exert considerable pressure on both the processing and memory subsystems. 

Quantization has emerged as a promising model compression method where parameters and/or activations are replaced with low-precision, quantized, fixed-point values. Despite such a transformation, quantized models can match the accuracy of full-precision models utilizing 32-bit floating-point (FP-32) while requiring fewer data transfers and storage space. 
Early works on binary neural networks (BNNs) \cite{DBLP:conf/nips/HubaraCSEB16} and XNOR-net \cite{DBLP:conf/eccv/RastegariORF16} demonstrated the potential advantages of extreme quantization, i.e., binarization.
BNNs are 1-bit quantized models where all weights and activations are represented by two values, -1/0 and +1, significantly decreasing the memory footprint.
Additionally, to speed up the inference, the multiplication/addition operations are switched out for less complex operations like the XNOR logical operation and bit count \cite{DBLP:conf/eccv/RastegariORF16}. 
However, this superior performance is achieved at the cost of a significant accuracy drop in deep neural networks.

To address the issue of significant accuracy loss, some prior work \cite{DBLP:conf/eccv/LiuWLYLC18, DBLP:conf/iclr/MartinezYBT20, DBLP:conf/fccm/NazemiFEKSP21, DBLP:conf/eccv/LiuSSC20, DBLP:conf/fpga/ZhangPLCCZ21} propose to modify the well-known architectures (e.g., ResNet) as they show the network architecture can affect BNN performance.
Although some of these works can achieve improved accuracy, their proposed models cannot be efficiently deployed on hardware platforms. 
For instance, ReActNet \cite{DBLP:conf/eccv/LiuSSC20} significantly improves the accuracy by activation shifting and reshaping the MobileNet V1 \cite{DBLP:journals/corr/HowardZCKWWAA17} architecture at the cost of increasing the parameters such that the total number of parameters is about 30 million more than the number of parameters in MobileNet V2 \cite{DBLP:conf/cvpr/SandlerHZZC18}.

This paper presents a novel building block called Blend module, which utilizes binary convolution on its main path and fixed-point convolution on its skip path. 
The key contributions of this work can be summarized as follows.
 \begin{itemize}[leftmargin=10pt]
        \vspace{-1.5mm}
        \item  We present NeuroBlend, a hardware-friendly neural network architecture with binary activations, where all convolutional layers are computed using binary multiply-accumulate (BMAC) operations (except on skip paths that utilize fixed-point convolutions). On a ResNet-20-like architecture designed for the CIFAR-10 dataset, NeuroBlend outperforms the state-of-the-art binary-based implementation \cite{DBLP:conf/fpga/ZhangPLCCZ21} by 0.8\% in top-1 accuracy.
        \item We introduce a powerful and flexible compiler for mapping a given NeuroBlend inference engine running on any dataset onto our optimized accelerator design by converting the network model to a computational graph, scheduling the graph's execution, and optimizing its nodes by leveraging intrinsic fusions of the convolution and batch normalization layers.
        \item We present a flexible FPGA-based design that enables the DSP-based realization of BMAC operations. The reconfigurability of DSPs for performing bitwise logic operations is utilized to achieve a low-power implementation.
        \item We apply our transformations, integrate our blocks to MLPMixers, and improve the naively binarized MLPMixer models by ~6\%. To the best of our knowledge, this is the first paper that presents a binary model of MLPMixers with negligible accuracy drop.
\end{itemize}

\section{Preliminaries}
\label{sec:prelim}

\begin{figure}[t]
     \centering
     \begin{subfigure}[b]{0.25\textwidth}
         \centering
         \includegraphics[width=\textwidth]{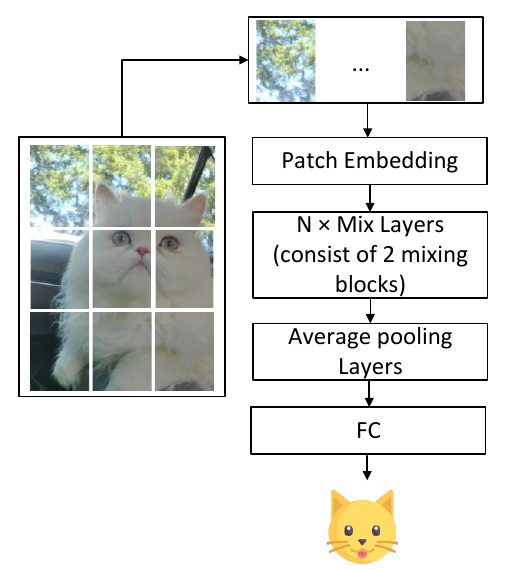}
         \caption{}
         \label{fig:mlpmixer1}
         \vspace{-3mm}
     \end{subfigure}
     \hfill
     \begin{subfigure}[b]{0.18
     \textwidth}
         \centering
         \includegraphics[width=\textwidth]{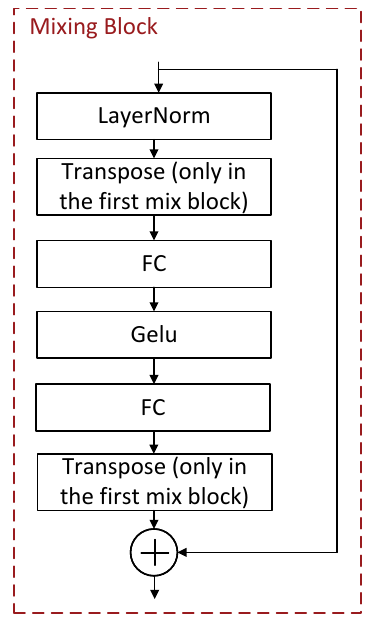}
         \caption{}
         \label{fig:mlpmixer2}
         \vspace{-3mm}
     \end{subfigure}
     \vspace{-2mm}
        \caption{\small Overview of MLPMixer model. (a) Overall MLPMixer and (b) mixing block architectures.}
        \label{fig:mlpmixer}
    \vspace{-6mm}
\end{figure}

In this section, we briefly introduce the MLPMixer, the conventional BNN model, and discuss some recent BNN architectures that have significantly improved accuracy.

\subsection{MLPMixer}
\label{subsec:mlmixer_back}
To further reduce the inductive biases introduced by CNNs, MLP-Mixer has recently proposed a more straightforward solution that is fully based on multi-layer perceptrons (MLPs) \cite{DBLP:conf/nips/TolstikhinHKBZU21} (see fig.~\ref{fig:mlpmixer}). 
The basic layer in MLP-Mixer consists of two components: the channel-mixing block and the token-mixing block. Each of these mixing blocks has the same units and is shown in Fig. \ref{fig:mlpmixer2}. In the channel-mixing block, the feature map is projected along the channel dimension for the communications among various channels, while the feature map is projected along the spatial dimension, and communications among spatial locations are accomplished concurrently by the token-mixing block.

\subsection{Conventional BNN}
\label{subsec:conv_bnn}

BNNs have several properties that enable more efficient mapping to FPGAs without affecting network accuracy. For implementing conventional BNN models that are binarized (both weights and activations are quantized to 1-bit values), the product of a binary weight and activation can be replaced with a binary XNOR operation.

Furthermore, by assuming that an unset bit represents -1 and a set bit represents +1, there are only two possible values of +1 and -1 for the result of the XNOR operation and, thus, synapse input. Therefore, the summation of a binary dot product can be implemented by a popcount operation that counts the number of set bits instead of accumulation with signed arithmetic. 
Furthermore, all BNN layers use batch normalization on convolutional or fully connected layer outputs and then apply the sign function to determine the output activation. For hardware implementation, the same output activation can be computed via thresholding \cite{DBLP:conf/fpga/UmurogluFGBLJV17}. Lastly, Max-pooling on binary activations can be implemented in hardware using the Boolean OR function.  

\subsection{Prior Work on State-of-the-art BNN Architectures}
\label{subsec:prior_work}
The state-of-the-art BNN models achieved a high accuracy using blocks similar to ResNet models \cite{DBLP:conf/cvpr/HeZRS16}. 
Martinez et al. \cite{DBLP:conf/iclr/MartinezYBT20} presented a strong baseline for BNN models which is based on the modified ResNet block suitable for 1-bit CNNs. 
They used double skip connections and PReLU (PReLU was first introduced in \cite{DBLP:journals/corr/abs-1904-05868}) as the activation function. They follow the idea of using real-valued downsample layers proposed in \cite{DBLP:conf/eccv/LiuWLYLC18} that improves the accuracy significantly.
%
%
%
They also presented a method wherein the architectural gap between real and binary networks is, step by step, bridged via a sequence of teacher-student pairs.

More recently, a more accurate BNN model is proposed called ReActNet \cite{DBLP:conf/eccv/LiuSSC20} to mitigate the precision gap between the binarized model and its counterpart of real-valued. ReActNet took one step further and is based on MobileNetV1 \cite{DBLP:journals/corr/HowardZCKWWAA17} architecture. It reaches a top-1 accuracy of 69.4\% in the IMAGENET data set using 4.82G BMACS with a 4.6 MB model size. 
The key block in ReActNet is a biased PReLU (BPReLU) activation function that shifts and reshapes the feature maps between two convolutional layers. This substantially improves the accuracy of the model. 
%

To improve ReActNet \cite{DBLP:conf/eccv/LiuSSC20}, The authors in \cite{DBLP:conf/fpga/ZhangPLCCZ21} have proposed FracBNN which employs a dual-precision activation scheme to compute features with up to two bits, using an additional sparse binary convolution layer. They have achieved MobileNetV2-level accuracy with competitive model size.

\section{The Proposed Building Blocks}
\label{sec:bnn}

\begin{figure}[t]
     \centering
     \begin{subfigure}[b]{0.16\textwidth}
         \centering
         \includegraphics[width=\textwidth]{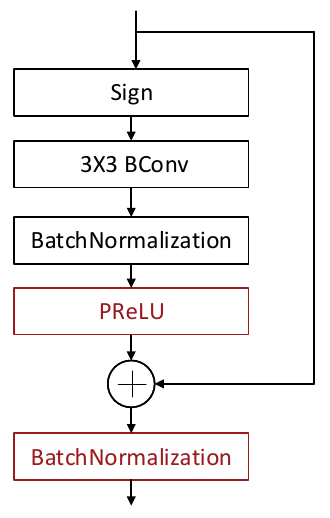}
         \caption{Normal block}
         \label{fig:buiding_blocks1}
         \vspace{-2mm}
     \end{subfigure}
     \hfill
     \begin{subfigure}[b]{0.28\textwidth}
         \centering
         \includegraphics[width=\textwidth]{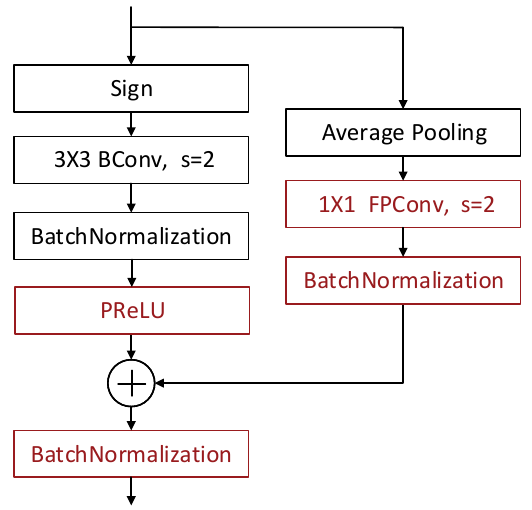}
         \caption{Downsample block}
         \label{fig:buiding_blocks2}
         \vspace{-2mm}
     \end{subfigure}
     \vspace{-2mm}
        \caption{\small The proposed building blocks. The differences with respect to real-to-binary blocks are highlighted in red.}
        \label{fig:buiding_blocks}
        \vspace{-6mm}
\end{figure}

The proposed BNN model comprises two types of building blocks, as shown in Fig.~\ref{fig:buiding_blocks}: one with no operations on the skip path and another with average pooling, convolution, and batch normalization on its skip path.
There are a few differences between the presented building blocks compared to prior work.

First, both types of building blocks include a batch normalization layer as their final output layer.
This batch normalization layer, which does not include any data-driven, trainable channel rescaling/shifting parameters, ensures the output of each block is in a range centered around zero, and is, therefore, amenable to fixed-point or binary quantization.
Adding channel rescaling/shifting to this batch normalization layer tends to reduce the classification accuracy by a few percentage points.

Second, a PReLU activation is placed in the main path and before the addition operation, which yields improvements in classification accuracy compared to other activation functions such as ReLU \cite{DBLP:conf/iclr/MartinezYBT20}.

\begin{figure}[t]
         \centering
         \includegraphics[width=0.40\columnwidth]{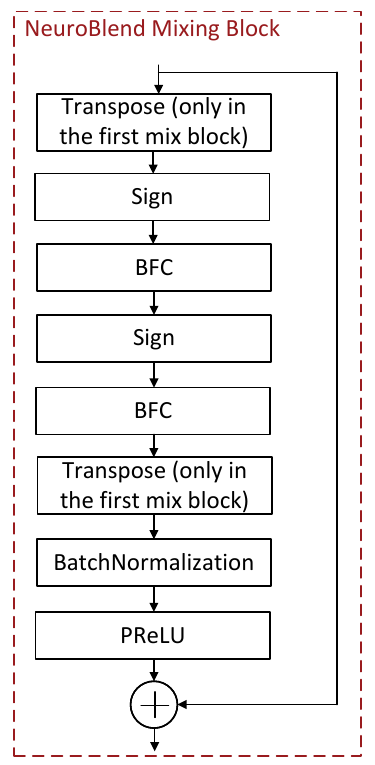}
         \vspace{-3mm}
        \caption{\small The proposed Mixing block in this paper.}
        \label{fig:blendmlpmixer}
        \vspace{-5mm}
\end{figure}

Last but not least, the building blocks are designed with the hardware implementation cost in mind. For example, as shown in Fig.~\ref{fig:comp_opt}, the final batch normalization layer of a building block can be fused into the next building block, resulting in a thresholding operation in the main path and modified convolutional weights in the skip path (see details in Sec.~\ref{subsec:comp-opt}).
A similar fusing can be performed for batch normalization in the skip path, all leading to reduced end-to-end inference latency.
We extend our building blocks to the MLPMixer model and propose a new Mixing block, as shown in Fig. \ref{fig:blendmlpmixer}. 

\section{Proposed Accelerator Design}
\label{sec:acc-des}
In this section, we describe the proposed hardware accelerator and the associated compiler to perform optimization tailored to the employed accelerator design. 
\subsection{Compiler Optimization}
\label{subsec:comp-opt}
Our compiler performs optimizations tailored to the employed accelerator design and fuses operations to reduce the hardware cost and generate an intermediate graph. Finally, our compiler compiles the intermediate graph, extracts the required parameters for the accelerator design, and generates a static schedule. 

\begin{figure*}[tb]
     \centering
     \begin{subfigure}[b]{0.32\textwidth}
         \centering
         \includegraphics[width=\textwidth]{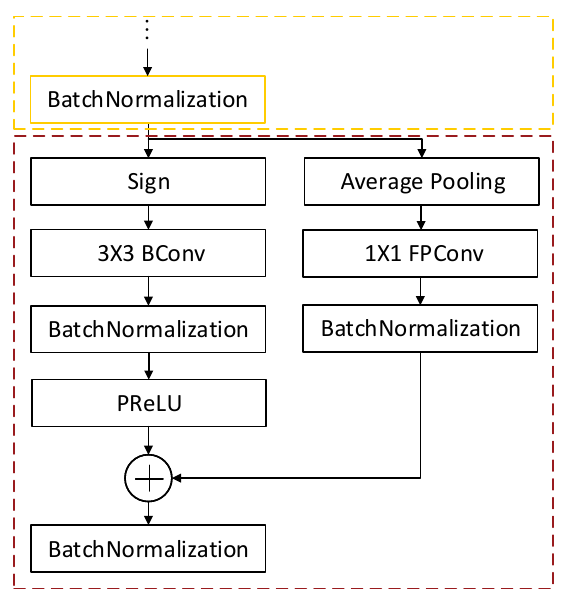}
         \caption{}
         \label{fig:compt_opt1}
         \vspace{-3mm}
     \end{subfigure}
     \hfill
     \begin{subfigure}[b]{0.32\textwidth}
         \centering
         \includegraphics[width=\textwidth]{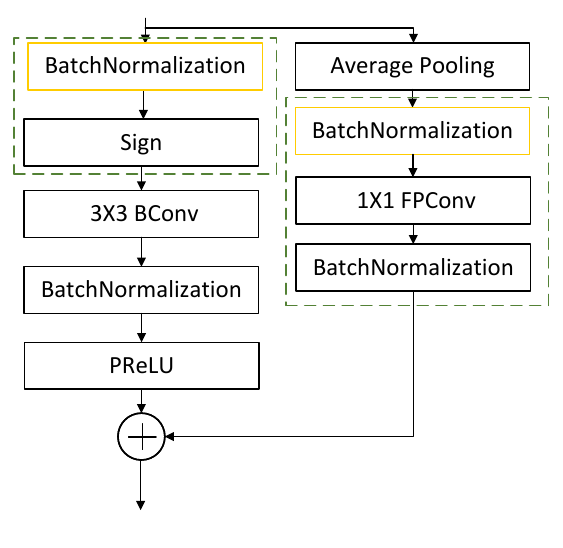}
         \caption{}
         \label{fig:compt_opt2}
         \vspace{-3mm}
     \end{subfigure}
     \hfill
     \begin{subfigure}[b]{0.32\textwidth}
         \centering
         \includegraphics[width=\textwidth]{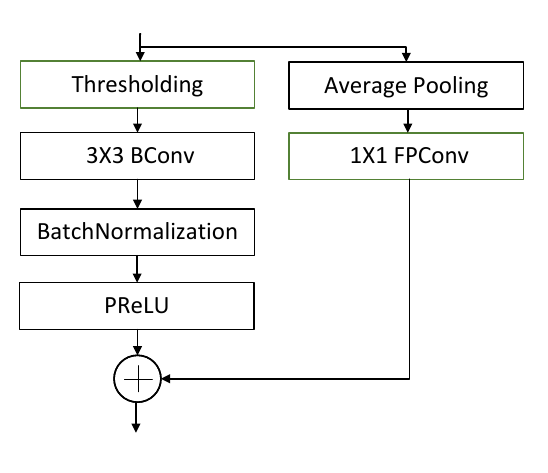}
         \caption{}
         \label{fig:compt_opt3}
         \vspace{-3mm}
     \end{subfigure}
     \vspace{-2mm}
        \caption{\small (a) The proposed building block. (b) Distribute the BN from the precedent NeuroBlend block to both main and skip paths. (c) Merge components that result in the optimized NeuroBlend block. Note that the red and yellow dash boxes show the precedent and current building blocks. The green dash boxes show the potential components for merging.}
        \label{fig:comp_opt}
        \vspace{-5mm}
\end{figure*}

All BNN layers proposed in this paper (cf. Fig~\ref{fig:buiding_blocks}) begin with a sign function (SG) and end with a batch normalization (BN) operation. %
First, we move the last BN block of a layer to the next layer and pass it through both feed-forward and skip paths of the next layer (cf. Fig.~\ref{fig:compt_opt2}. 
Moreover, the BN block that comes before the SG block can be replaced by a thresholding (TH) function in order to reduce the hardware cost (c.f. Fig. \ref{fig:compt_opt3}). 
Using such a technique, we can process the input feature map using an unsigned comparison and avoid expensive operations such as multiplication that are required in the BN block. 
Reference \cite{DBLP:conf/fpga/UmurogluFGBLJV17} explains how the hardware cost of a regular BN-SG block is reduced from 2 DSPs, 55 FFs, and 40 LUTs for separate BN and SG computations to merely 6 LUTs for the TH block computations using such a technique.

In addition, the BN block that is passed onto the skip path is also fused with the CONV layer in the skip path if one exists. In summary, we will replace a BN-CONV-BN sequence of layers on the skip path with a CONV layer (c.f. Fig. \ref{fig:compt_opt3}). More details of such fusion are provided below. Note that the BN block that is passed onto the skip path is passed through the average pooling block in the inference (c.f. Fig. \ref{fig:compt_opt2}).
This simplification also reduces the chances of encountering overflow/underflow even if we assign fewer bits than required bits for summation because the weights/biases of the fused BN-CONV-BN layer will help normalize this layer's outputs. In other words, the output of BMAC can be considered 16 bits, and no quantization is required.  

Algorithm \ref{alg:BN} shows the well-known batch normalization algorithm where the two parameters \(\gamma\) and \(\beta\) are learned during the training process. 
Note that \(\epsilon\) is a small constant value used to ensure that division-by-zero error is not encountered.

\begin{algorithm}[t]
  \small
  \caption{\small Batch normalization for activation $y$ in a mini-batch, $y'$ is the normalized result}
  \label{alg:BN}
   \hspace*{\algorithmicindent} \textbf{Input}  \(\mathcal{B}\) = \{\(y_0, y_1,..., y_m\)\}; \(\gamma\); \(\beta\)\\
   \hspace*{\algorithmicindent} \textbf{Output}  \(y'_i\) = $BN_{\gamma,\beta}(y_i)$
  \begin{algorithmic}[1]
    \STATE $\mu_{\mathcal{B}} \gets \frac{1}{m} \sum_{i=1}^{m}y_i$ \COMMENT{mini-batch mean}
    \STATE $\sigma_{\mathcal{B}}^2 \gets \frac{1}{m} \sum_{i=1}^{m}(y_i-\mu_{\mathcal{B}})^2$ \COMMENT{mini-batch variance}
    \STATE $\hat{y_i} \gets \frac{y_i-\mu_{\mathcal{B}}}{\sqrt{\sigma_{\mathcal{B}}^2+\epsilon}}$ \COMMENT{normalize}
    \STATE $y'_i \gets \gamma \hat{y_i} + \beta \equiv BN_{\gamma,\beta}(y_i)$ \COMMENT{scale and shift}
\end{algorithmic}
\end{algorithm}

When the BN layer is fused into a subsequent convolutional layer, the fused layer's computations may be written as,
\begin{equation}
\label{equ:bn-conv}
\small
    y' =  \mathbold{w} (\gamma \frac{ \mathbold{x} - \mathbold{\mu'}_{\mathcal{B}}}{\sqrt{\mathbold{\sigma'}_{\mathcal{B}}^2+\epsilon}} + \mathbold{\beta'}) + \mathbold{b}
\end{equation}
%
Hence, the new fused parameters can be calculated as:
\begin{equation}
\small
\begin{gathered}
    \mathbold{w'} = \frac{\mathbold{\gamma'} \mathbold{w}}{\sqrt{\mathbold{\sigma'}_{\mathcal{B}}^2+\epsilon}}, ~~~ \mathbold{b'} = -\frac{\mathbold{\gamma'} \mathbold{w}\mathbold{\mu'}_{\mathcal{B}}}{\sqrt{\mathbold{\sigma'}_{\mathcal{B}}^2+\epsilon}}  
    + \mathbold{w}\mathbold{\beta'} + \mathbold{b}
    \end{gathered}
\end{equation}
Note that $\mathbold{x}$ in Eqn. \ref{equ:bn-conv} is the output of previous layer, so the 4-D $\mathbold{w}$ must be squeezed into a 2-D vector of size (\(c_{\mathrm{in}}\), \(c_{\mathrm{out}}\) $\times$ \(w_{\mathrm{k}}\) $\times$ \(h_{\mathrm{k}}\)). Indeed, if we use zero padding in convolutions, we will have 0 values entering the convolution after the BN. When we fuse the parameters, this must still be the case. So, to correctly fold $\mathbold{w}$ into the 2-D vector, we have to replace the default 0 values for padding by $\mathbold{\mu'}_{\mathcal{B}} -\frac{\mathbold{\beta'} \sqrt{\mathbold{\sigma'}_{\mathcal{B}}^2+\epsilon}}{\mathbold{\gamma'}}$, that is, we must apply the BN transformation to the padding. Hence, by multiplying the fused weights $\mathbold{w'}$ to the padding values, we apply the inverse of the BN transformation.

Using algorithm \ref{alg:BN} and proceeding with fused parameters as in the previous case, the weights and bias of the resulting fused BN-CONV-BN block may be expressed as:

\begin{equation}
\small
\begin{gathered}
    \mathbold{w''} = \frac{\mathbold{\gamma''} \mathbold{w'}}{\sqrt{\mathbold{\sigma''}_{\mathcal{B}}^2+\epsilon}}, ~~~ \mathbold{b''} = \mathbold{\beta''} + \mathbold{\gamma''} \frac{\mathbold{b'} - \mathbold{\mu''}_{\mathcal{B}}}{\sqrt{\mathbold{\sigma''}_{\mathcal{B}}^2+\epsilon}}
    \end{gathered}
\end{equation}

\subsection{Accelerator Architecture}
\label{subsec:acc-arc}
We adopted a heterogeneous streaming architecture where each layer uses its own hardware resources for this work. 
%

Our Accelerator consists of the following types of operations: 
1) $3 \times 3$ BNN convolution, 2) $1 \times 1$ Fixed-Point convolution, 3) Average pooling, 4) Max Pooling, 5) Linear layer, 6) Thresholding, 7) BN-PReLU/BN-PReLU-BN, and 8) Residual connection and summation.

%
We have designed and implemented an efficient accelerator that supports these operations on FPGA devices. We separate our design into three domains: binary, fixed-point, and joint domains. The BNN (FPNN) convolutions are placed in binary (fixed-point) domains, whereas all other operations are in the joint domain. 
Our hardware is designed by using Xilinx's High-Level Synthesis (HLS) tools. 
In the following, we will describe the hardware engines and the HLS design techniques in detail.
\vspace{-2mm}
\subsubsection{BNN and FPNN Block}
\label{subsubsec:bnn-fpnn}
The BMAC/FPMAC of the proposed accelerator are mapped to DSPs.
Each DSP block (i.e., DSP48E2 in Xilinx FPGA) is capable of performing a 18 by 27-bit or a 48-bit bitwise logic operation including AND, OR, NOT, NAND, NOR, XOR, and XNOR. 
We used this feature to perform 48 XNOR operations simultaneously as one BMAC operation. Note that similar 48 XNOR operations on LUTs result in the usage of 48 LUTs and 49 FFs.
So, the input feature map and weights must be packed into groups of size 48. For this purpose, We pack bits along the channel dimension into 48-bit unsigned integers for concurrent access. We will highlight the advantage of mapping logic operations to DSPs in Section \ref{subsec:res-hardware}. 

Our design results in a balanced usage of hardware resources since LUTs are mostly used for other operations while DSPs are used to do BMAC/FPMAC operations. 
Note that the reconfigurability of DSPs in each clock cycle can be used to switch between BMAC and FPMAC operations. 
This is very helpful in the case of a homogeneous single execution engine design where we only use a fixed set of resources for all layers, although this is not in the scope of the present paper. Note that option can be used in single execution engine architecture. In the streaming architecture, each unit has its own resource. A 2D array of DSPs with the size of 32 * 32 is designed to perform FPMAC operations. 
%
\vspace{-2mm}
\subsubsection{FP/B Joint Blocks}
\label{subsubsec:bnn-fpnn-joint}
In addition to the accelerators for the convolution operations, which account for the majority of the computations in a vision neural network, we also implement hardware accelerators for other operations that must be performed in the joint domain, including the max (average) pooling and thresholding blocks. For instance, the TH unit compares each output activation from the previous layer with a programmable threshold and outputs +1 (0) if the output activation is greater (smaller) than the corresponding threshold. Since these TH blocks only contain channel-wise parameters, their impact on the total number of model parameters is negligible. Although we can achieve the maximum concurrency by processing all output activations simultaneously, this would require a lot of resources, including both memories (we must increase the number of access lines by partitioning the data to several memory blocks) and TH block. The performance gain may not be worth the cost of additional resources needed. We use the greatest common divisor (GCD) of the height of the systolic array in the FPNN block (e.g., 32) and the width of the array of BMACs in the BNN block (e.g., 48) as the parallelism factor for all blocks in the joint domain. 

\section{Experimental Results}
\label{sec:exp-res}
For evaluation purposes, we targeted a high-end Virtex\textregistered ~UltraScale+ FPGA (Xilinx VU9P FPGA, which is available in the cloud as the AWS EC2 F1 instance). 
This FPGA contains approximately 2.5 million logic elements and approximately 6,800 DSP units.
We use Vitis 2020.2 for mapping to the FPGA and set the target frequency to 340 Mhz. We also use the Vivado power report provided by Xilinx to assess the power consumption of each design. 
Finally, we evaluate our proposed method on a well-known CNN, i.e., ResNet-20 \cite{DBLP:conf/cvpr/HeZRS16} and MLPMixers \cite{DBLP:conf/nips/TolstikhinHKBZU21} and a commonly used computer-vision dataset for object recognition, i.e., the CIFAR-10 \cite{krizhevsky2009learning} dataset.

\subsection{Experimental setup}
\label{subsec:exp-set}
In the case of MLPMixers, the resolution of the input image is 32*32, and the patch size that the experiments are based on is 4*4. So, we have S non-overlapping image patches that are mapped to a hidden dimension $C$. $D_{S}$ and $D_{C}$ are tunable hidden widths in the token-mixing and channel-mixing MLPs, respectively. The summary of design specifications is shown in Table \ref{tab:mlpmixerspec}.

\begin{table}[tb]
   \caption{Summary of design specifications for MLPMixers used in this paper. The "S" and "B" (small and base)
models scales down follow Tolstikhin et al. \cite{DBLP:conf/nips/TolstikhinHKBZU21}. The notation "B/4" means the model of base scale with patches of resolution 4*4.} 
\vspace{-2mm}
   \label{tab:mlpmixerspec}
   \small 
   \centering 
   \begin{tabular}{|l|c|c|c|c|} 
   \toprule[\heavyrulewidth]
   \textbf{Specification} & S/4 & B/4  & 2S/4 \\
   \midrule
   \textbf{Sequence length S} & 64 & 64 & 64 \\
   \midrule
   \textbf{Hidden size C} & 128 & 192 & 256 \\
   \midrule
   \textbf{Patch resolution} & 4*4 & 4*4 & 4*4 \\
   \midrule
   \textbf{MLP dimension $D_{C}$} & 512 & 768 & 1024  \\
   \midrule
   \textbf{MLP dimension $D_{S}$} & 64 & 96 & 128  \\
   \midrule
   \textbf{Number of layers} & 8 & 12 & 8 \\
   \bottomrule[\heavyrulewidth] 
   \end{tabular}
   \vspace{-4mm}
\end{table}

\subsection{ResNet-18 and ResNet-20}
Table~\ref{tab:ResNet-20} demonstrates the superiority of NeuroBlend on ResNet-20 and ResNet-18 compared to other binary-based approaches on CIFAR-10 and CIFAR-100 datasets. 
We can observe that NeuroBlend achieves a higher top-1 accuracy by 0.8\% for the CIFAR-10 dataset and 1.33\% for the CIFAR-100 dataset compared to the state-of-the-art BNNs and improves the accuracy by 1.5\% to about 9\% compared to the other approaches. NeuroBlend-20 achieves the same model size as FracBNN even while keeping the first and last layer full precision (i.e., 16-bit fixed-point).

\begin{table}[tb!]
   \caption{Classification accuracy of ResNet-20-like and Resnet-18 models on CIFAR-10 and CIFAR-100, respectively.} 
   \vspace{-2mm}
   \label{tab:ResNet-20}
   \centering%
   \resizebox{\columnwidth}{!}{%
   \begin{tabular}{l c | l c}
   \toprule[\heavyrulewidth]
   \multicolumn{2}{c|}{CIFAR-10} & \multicolumn{2}{c}{CIFAR-100}  \\ \midrule
   Approach                                         & Accuracy (\%) & Approach                                         & Accuracy (\%) \\
   \midrule
   DoReFa-Net \cite{DBLP:journals/corr/ZhouNZWWZ16} & 79.3 & Bi-RealNet-18 \cite{DBLP:conf/eccv/LiuWLYLC18} 
   & 63.51           \\
   DSQ \cite{DBLP:conf/iccv/GongLJLHLYY19}          & 84.1 & ReActNet-18 \cite{DBLP:conf/eccv/LiuSSC20}        & 70.76           \\
   ReActNet \cite{DBLP:conf/eccv/LiuSSC20}          & 85.8  &   AresB-Net-18\cite{kim2021aresb} 
   & 71.98       \\
   IR-Net \cite{DBLP:conf/cvpr/QinGLSWYS20}         & 86.5  & PresB-Net-18\cite{shin2022presb}
   & 72.30        \\
   FracBNN \cite{DBLP:conf/fpga/ZhangPLCCZ21}       & 87.2 & Hyper-BinaryNet\cite{wang2021gradient} 
   & 72.40          \\
   \midrule
   \textbf{NeuroBlend-20}                                & \textbf{88.0} & \textbf{NeuroBlend-18}                                & \textbf{73.73} \\
   \bottomrule[\heavyrulewidth] 
   \end{tabular}}
   \vspace{-5mm}
\end{table}

\begin{table}[tb!]
  \caption{\small Comparison between the hardware metrics of different types of NeuroBlend-20 implementation on CIFAR-10.}
  \vspace{-2mm}
  \label{table:resource-util}
  \resizebox{\columnwidth}{!}{%
  \centering 
  \begin{tabular}{|c|c|c|c|c|c|c|} 
  \toprule[\heavyrulewidth]
   \textbf{Approach}  & DSP48E2 & LUT & FF & BRAM\_18K &  \shortstack{$f_{target}$} & Power \\ 
  \midrule
  \textbf{\shortstack{DSP-based \\ NeuroBlend-20}} & \shortstack{2240 \\ (32\%)} &  \shortstack{10K \\ (1\%)} & \shortstack{30K\\ (1\%)} & \shortstack{912\\ (42\%)} &  \shortstack{342 \\ MHz} & \shortstack{6.1 \\ w}\\ 
  \midrule
   \textbf{\shortstack{LUT-based \\ NeuroBlend-20}} & \shortstack{1132 \\ (17\%)} &  \shortstack{500K \\ (50\%)}  & \shortstack{900K\\ (34\%)} & \shortstack{912\\(42\%)} & \shortstack{342  \\ MHz} & \shortstack{15.3 \\ w} \\ 
  \midrule
  \end{tabular}}
  \vspace{-6mm}
\end{table}


\subsection{Hardware cost and performance of NeuroBlend}
\label{subsec:res-hardware}
In this section, we evaluate our NeuroBlend-20 model on the FPGA platform. Compared to the BNN accelerator in FracBNN \cite{DBLP:conf/fpga/ZhangPLCCZ21}, our design achieves a higher frame rate (3846 FPS vs. 2807 FPS reported in FracBNN \cite{DBLP:conf/fpga/ZhangPLCCZ21}) and higher working frequency (342 MHz vs. 250 MHz) while yielding higher accuracy. Note that the FPGA used in this paper is a server-class FPGA while authors in \cite{DBLP:conf/fpga/ZhangPLCCZ21} deployed their design on an embedding FPGA with fewer resources. However, the frame rate comparison is fair since they unroll the entire model, which is similar to what we achieved. The unrolling is feasible because ResNet-20-based models on CIFAR-10 are very compact and can be fitted into an FPGA to eliminate unnecessary transactions between the logic blocks and the DDR memory.

Table \ref{table:resource-util} compares the hardware cost of two approaches for implementing the NeuroBlend-20 on the CIFAR-10 dataset where the DSP-based NeuroBlend-20 is the approach presented in Section \ref{subsubsec:bnn-fpnn}. The LUT-based is a naive implementation where logic operations are performed using LUTs. Note that DSP usage in LUT-based implementation is for the first and last layers. Decreasing the LUT usage results in power saving. Our measurements show that the proposed implementation yields 2.5x lower power consumption.

\subsection{MLPMixers}
As table \ref{tab:mlpmixerResults} shows, our model (i.e., BlendMixer) outperforms the naively binarized version of MLPMixer (i.e., BinaryMixer) yet achieves a smaller memory footprint (i.e., model size). Increasing the model size (e.g., model B/4) reduces the accuracy drop due to binarization. 
Note that FPMAC operations in both BinaryMixer and BlendMixer are due to the patch embedding block and the last FC layer.
When it comes to analyzing the performance, BlendMixer can achieve 0.02ms (0.06ms) latency for a small (base) model, while MLPMixer can process an image in 0.24ms (1.14ms). 

\subsection{Ablation study with normalization unit in mixing blocks}
To further evaluate the influence of the normalization unit on the proposed Mixing blocks, we performed an ablation study with different normalization units, layer normalization as suggested in the paper, batch normalization over channels, and batch normalization over patches. As shown in Table \ref{tab:norm}, the Mixing block with batch normalization over channels outperforms other models. Using batch normalization also reduces the memory footprint, requiring storing fewer parameters. Note that the MLPMixer consists of the proposed mixing blocks called BlendMixer.

\subsection{Computation/memory cost vs accuracy in mixing blocks}
In this section, we assess the trade-off between computation/memory complexity and result accuracy. As demonstrated in Table \ref{tab:mlpmixerbestesults}, we can manipulate the type of computation in FC layers of mixing blocks to improve the accuracy at the cost of computation and memory complexity. The BlendMixer(BB/BB-2S/4) is simply the widened version of BlendMixer S/4, and BlendMixer(BB/FPB-2S/4) is the same as the previous architecture with the exception that the FC layer of the second mixing block being calculated in FP format. The other models are named accordingly. The BlendMixer(BB/FPB-2S/4) achieves a comparable accuracy (less than 1\% accuracy drop compared to MLPMixer S/4). 

\begin{table}[tb]
   \caption{ \small Comparison between the mapping metrics of BlendMixer with those of MLPMixer  \cite{DBLP:conf/nips/TolstikhinHKBZU21} and BinaryMixer on CIFAR-10 dataset.} 
   \vspace{-2mm}
   \label{tab:mlpmixerResults}
   \resizebox{\columnwidth}{!}{%
   \centering 
   \begin{tabular}{c|c|ccccc} 
   \toprule[\heavyrulewidth]\toprule[\heavyrulewidth] 
   \textbf{Model Spec} & \textbf{Model}  & \textbf{\shortstack{Precision \\ (W/A)}} & \textbf{\shortstack{Model Size \\ (MB)}} &  \textbf{\shortstack{FPMAC \\($\times 10 ^8$)}} & \textbf{\shortstack{BMAC \\($\times 10 ^8$)}} & \textbf{\shortstack{ Top-1 \\ (\%)}}  \\
   \midrule
   \multirow{3}{*}[-0.4ex]{\rotatebox[origin=c]{90}{\textbf{S/4}}}
   & MLPMixer     & 16/16 & 2.5 & 0.76 & 0 & 92.25\\
   & BinaryMixer  & 1/1 & 0.41 & 0.005 & 0.75 & 74.38 \\
   & \textbf{BlendMixer}   & 1/1 & \textbf{0.15} & 0.005 & 0.75 & \textbf{80.43} \\
   \midrule
   \multirow{3}{*}[-0.4ex]{\rotatebox[origin=c]{90}{\textbf{B/4}}} 
   & MLPMixer     & 16/16 & 8.41 & 3.83 & 0 & 92.93 \\
   & BinaryMixer  & 1/1 & 1.07 & 0.015 & 3.82 & 82.5 \\
   & \textbf{BlendMixer}   & 1/1 & \textbf{0.49} & 0.015 & 3.82 & \textbf{87.34} \\
   \midrule
   
   \bottomrule[\heavyrulewidth] 
   \end{tabular}}
   \vspace{-4mm}
\end{table}

\begin{table}[tb!]
   \caption{Accuracy comparison between BlendMixer S/4 with different normalization unit.} 
   \vspace{-2mm}
   \label{tab:norm}
   \resizebox{\columnwidth}{!}{%
   \centering 
   \begin{tabular}{|l|c|c|c|c|c|} 
   \toprule[\heavyrulewidth]
   \textbf{Normalization unit} & LayerNormalization & \shortstack{BatchNormalization \\ (over channels)} & \shortstack{BatchNormalization \\(over patches)} \\
   \midrule
   \textbf{Accuracy} & 90.60 & 91.96 & 90.57 \\

   \bottomrule[\heavyrulewidth] 
   \end{tabular}}
   \vspace{-6mm}
\end{table}

\begin{table}[tbh]
   \caption{ \small Comparison between different design choices for trading-off computation cost for accuracy. Each layer in BlendMixer has two mixing blocks, each of which contains two FCs. Symbols B and FP show the implementation of FCs where B (FP) is binary (fixed-point) implementation.} 
   \vspace{-2mm}
   \label{tab:mlpmixerbestesults}
   \resizebox{\columnwidth}{!}{%
   \centering 
   \begin{tabular}{c|ccccc} 
   \toprule[\heavyrulewidth]\toprule[\heavyrulewidth] 
   \textbf{Model}  & \textbf{\shortstack{Model Size \\ (MB)}} &  \textbf{\shortstack{FPMAC \\($\times 10 ^8$)}} & \textbf{\shortstack{BMAC \\($\times 10 ^8$)}} & \textbf{\shortstack{ Top-1 \\ (\%)}}  \\
   \midrule

   BlendMixer(BB/BB-2S/4)   & 0.58 & 0.021 & 6.03 & 88.39  \\
   BlendMixer(BB/FPB-2S/4)   & 2.42 & 0.063 & 3.36 & 91.35  \\
   BlendMixer(BB/BFP-2S/4)   & 2.42 & 0.063 & 3.36 & 91.28  \\
   BlendMixer(FPB/BB-2S/4)   & 0.70 & 0.023 & 5.7 & 90.66  \\
   BlendMixer(BFP/BB-2S/4)   & 0.70 & 0.023 & 5.7 & 90.53  \\
   \midrule
   \bottomrule[\heavyrulewidth] 
   \end{tabular}}
   \vspace{-5mm}
\end{table}

\section{Conclusion}
\label{sec:conclusion}
This paper introduces NeuroBlend, an innovative neural network design that makes use of the Blend module, a new building block that performs binary and fixed-point convolutions in the main and skip paths, respectively. Batch normalizations are deployed intelligently on both the main and skip paths within the Blend module as well as between adjacent Blend modules. NeuroBlend-20, a descendant of ResNet-20 trained on the CIFAR-10 dataset, achieves 88.0\% classification accuracy (0.8 percent better than the state-of-the-art binary neural network), yet 1.4x higher throughput. 
Finally, BlendMixer, our new block inspired by MLPMixer blocks, outperforms the naively binarized version of MLPMixer yet achieves a smaller memory footprint.




\bibliographystyle{ACM-Reference-Format}
\bibliography{sample-base}

\end{document}